\documentclass[reprint,amsmath,amssymb,aps]{revtex4-2}
\usepackage{graphicx}
\usepackage{dcolumn}
\usepackage{bm}
\usepackage{float}
\usepackage{color}
\usepackage[dvipsnames]{xcolor}
\usepackage{hyperref}
\usepackage{graphicx}
\usepackage{braket}
\usepackage{appendix}
\usepackage{chemmacros}
\usepackage{natbib}
\usepackage{amsmath}
\hypersetup{colorlinks=true, citecolor=blue, urlcolor=blue, linkcolor=blue}

\begin{document}

\preprint{APS/123-QED}

\title{Formation of spin-orbital entangled 2D electron gas in layer delta-doped bilayer iridate La$_{\delta}$Sr$_3$Ir$_2$O$_7$}

\author{Amit Chauhan}
\email{amitchauhan453@gmail.com}
\author{Arijit Mandal}
\author{B. R. K. Nanda} 
\email{nandab@iitm.ac.in}
\affiliation{$^1$Condensed Matter Theory and Computational Lab, Department of Physics, IIT Madras, Chennai-36, India}
\affiliation{$^2$Center for Atomistic Modelling and Materials Design, IIT Madras, Chennai-36, India}
\date{\today}
\begin{abstract}
5$d$ transition metal oxides host a variety of exotic phases due to the comparable strength of Coulomb repulsion and spin-orbit coupling. Herein, by pursuing density-functional studies on a delta-doped quasi-two-dimensional iridate Sr$_3$Ir$_2$O$_7$, where a single SrO layer is replaced by LaO layer, we predict the formation of a spin-orbital entangled two-dimensional electron gas (2DEG) which is sharply confined on two IrO$_2$ layers close to the LaO layer. In this bilayer crystal structure, an existing potential well is further augmented with the inclusion of positively charged LaO layer which results in confining the extra valence electron made available by the La$^{3+}$ ion. The confined electron is bound along crystal $a$ direction and is highly mobile in the $bc$ plane. A tight-binding model Hamiltonian involving the Ir-t$_{2g}$ orbitals is formulated to provide further insight into the formation of 2DEG. From the band structure point of view, now the existing half-filled $J_{eff}$ = 1/2 states are further electron doped to destroy the antiferromagnetic Mott insulating state of IrO$_2$ layers near to the delta-doped layer. This leads to partially occupied Ir upper-Hubbard subbands which host the spin-orbital entangled 2DEG. The IrO$_2$ layers far away from the interface remain insulating and preserve the collinear G-type magnetic ordering of pristine Sr$_3$Ir$_2$O$_7$. The conductivity tensors calculated using semi-classical Boltzmann theory at room temperature reveal that the 2DEG exhibits large electrical conductivity of the order of 10$^{19}$.  \end{abstract}
\maketitle

\section{Introduction}
The state-of-the-art growth techniques have paved the way to introduce dopants that can be confined in a single atomically thin layer, the so-called delta-doping ($\delta$-doping)   \cite{Wood1980,Chang2022,Liu1990}. $\delta$-doping is widely used to manipulate structures for fundamental importance as well as for novel applications \cite{Schubert1990}. Experimental realization of novel structures based on delta-doping \cite{Liu1990,Kim1993,KIM1995,Oubram2009,Chang2022} has been achieved by employing techniques such as molecular beam epitaxy \cite{Liu1990}, metalorganic chemical vapor deposition \cite{Kim1993}, flash lamp annealing \cite{Chang2022}, etc. Though the $\delta$-doping technique is widely applied in conventional semiconductors such as GaAs, it also paves the way to realize novel quantum phases in transition metal oxides and their heterostructures which involve chemically active $d$-electrons. For example, $\delta$-doped SrTiO$_3$ exhibits quantum Hall effect \cite{Matsubara2016} and Shubnikov-de Haas oscillations \cite{Jalan2010} arising from a two-dimensional electron gas (2DEG) which possess enhanced electron mobility \cite{Kozuka2010}.

\begin{figure}
    \centering  
   \includegraphics[angle=-0.0,origin=c,height=8cm,width=8.7cm]{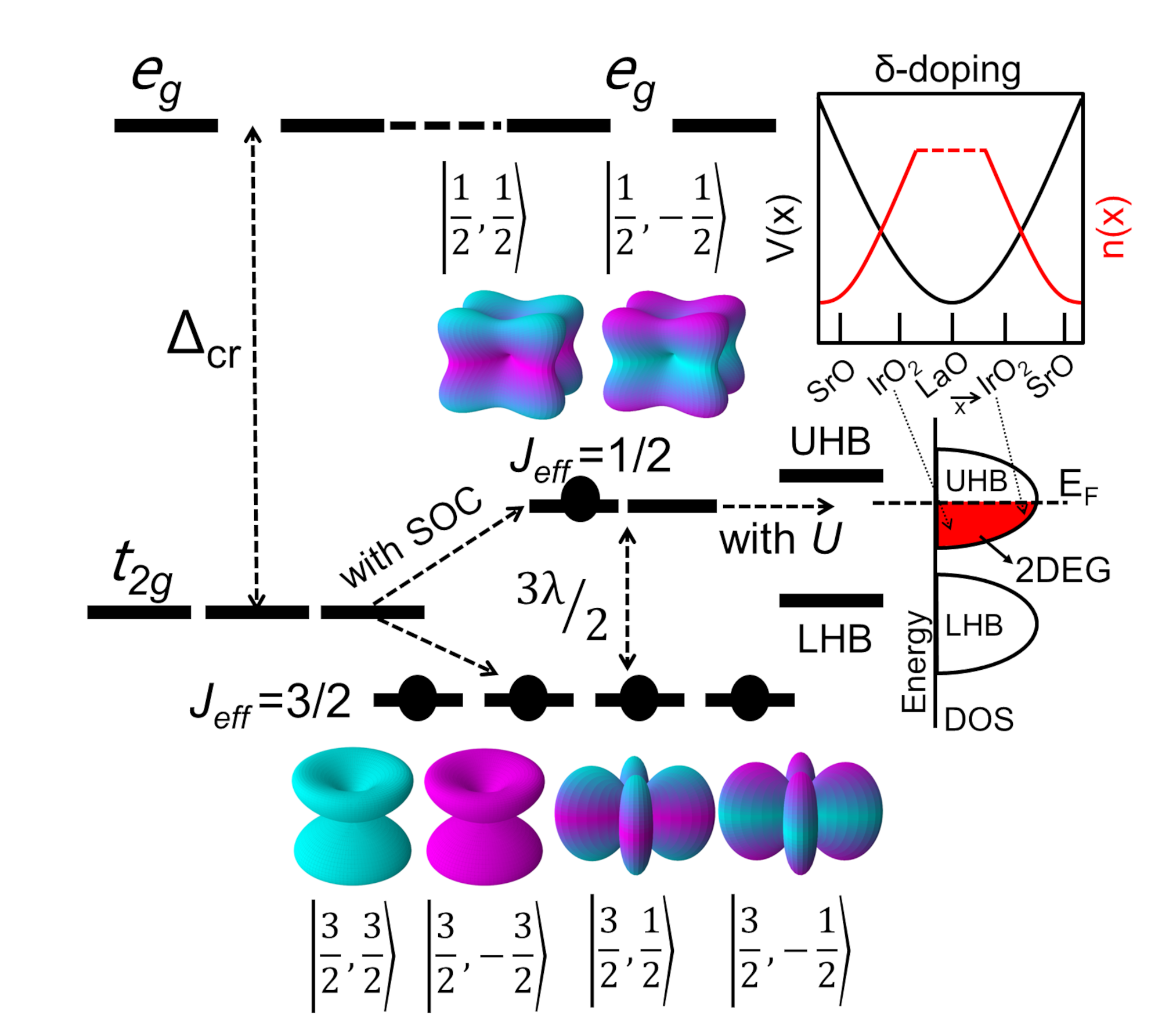}
    \caption{The octahedral crystal field splitting leading to the formation of three-fold degenerate $t_{2g}$ and two-fold degenerate $e_g$ manifolds. With strong SOC, while the $e_g$ states remain unperturbed, the $t_{2g}$ states further split to form four-fold and two-fold degenerate spin-orbital entangled $J_{eff}$ = 3/2 and $J_{eff}$ = 1/2 states. The aqua and magenta colors represent spin-up and spin-down weights, respectively. Further, the onsite Coulomb repulsion split the latter to form lower and upper Hubbard subbands. With $\delta$-doping
    , where a SrO layer is replaced by LaO layer, the extra La electron gets trapped in the potential well so that the UHB of IrO$_2$ layers, which are adjacent to LaO layer, gets partially occupied to host 2DEG.}
    \label{split-d-states}
\end{figure}

Recently, iridate heterostructures and interfaces have been epitaxially grown which facilitate the growth of $\delta$-doped iridate structures where an element is substituted with another element having excess holes or electrons. The 5$d$ quantum materials, specifically iridates, are extensively studied as they host exotic states such as spin liquid \cite{Okamoto2007,kenney2019,Takahashi2019}, Dirac and Weyl semimetals \cite{Chauhan2022,Ueda2018}, topological insulators \cite{Pesin2010}, etc., driven by competing interactions such as onsite Coulomb repulsion ($U$), spin-orbit coupling (SOC), Hund's coupling, etc. In Ruddlesden-Popper phases of strontium iridate, the Ir ions exhibit 4+ charge state and hence $d^5$ electronic configuration. Due to strong octahedral crystal field, the $d$ orbital degeneracy gets lifted, giving rise to $t_{2g}$ and $e_\mathrm{g}$ manifold. With the inclusion of strong SOC ($\approx$ 0.43 eV), the $t_{2g}$ manifold splits into spin-orbital entangled pseudo-spin $J_{eff}$ = 3/2 and $J_{eff}$ = 1/2 states (see Fig. \ref{split-d-states}). Due to Ir-5$d^5$ configuration, the $J_{eff}$ = 3/2 ($m_J$ = $\pm$ 1/2, $\pm$ 3/2) states are completely occupied, leaving a single hole in the $J_{eff}$ = 1/2 state \cite{Chauhan2022}. 
Further, with the inclusion of onsite correlation effect, these doubly degenerate $J_{eff}$ = 1/2 states further split into lower and upper Hubbard subbands (LHB and UHB) with a gap in between. While the former is occupied, the latter spin-orbital entangled electron state can host 2DEG with intriguing transport properties upon electron doping.

The Sr$_3$Ir$_2$O$_7$ (SIO-327, Ir-$d^5$), the quasi-two-dimensional member in Ruddlesden Popper phases of strontium iridate, has attracted a great deal of attention among theoreticians and experimenters alike, as it builds a perfect platform to realize novel properties via carrier doping of the half-filled Mott state. It has been established experimentally and theoretically that a dilute electron doping via La substitution causes collapse of both Mott and long-range G-type N\'eel state in (Sr$_{1-x}$La$_x$)$_3$Ir$_2$O$_7$ (x $\approx$ 0.04) \cite{Swift2018,Hogan2015}. Moreover, a recent angle-resolved photoemission spectroscopy (ARPES) and optical reflectivity measurements report negative electronic compressibility \cite{He2015} and charge density wave (CDW) instability \cite{Chu2017} in electron doped SIO-327. Apart from electron doping in SIO-327, a very recent resonant inelastic X-ray scattering (RIXS) study reports the formation of an excitonic insulating state in pristine SIO-327 which may have the potential to exhibit new functionalities upon electron doping \cite{Mazzone2022}.
\begin{figure*}
    \centering
    \includegraphics[angle=-0.0,origin=c,height=6.5cm,width=12cm]{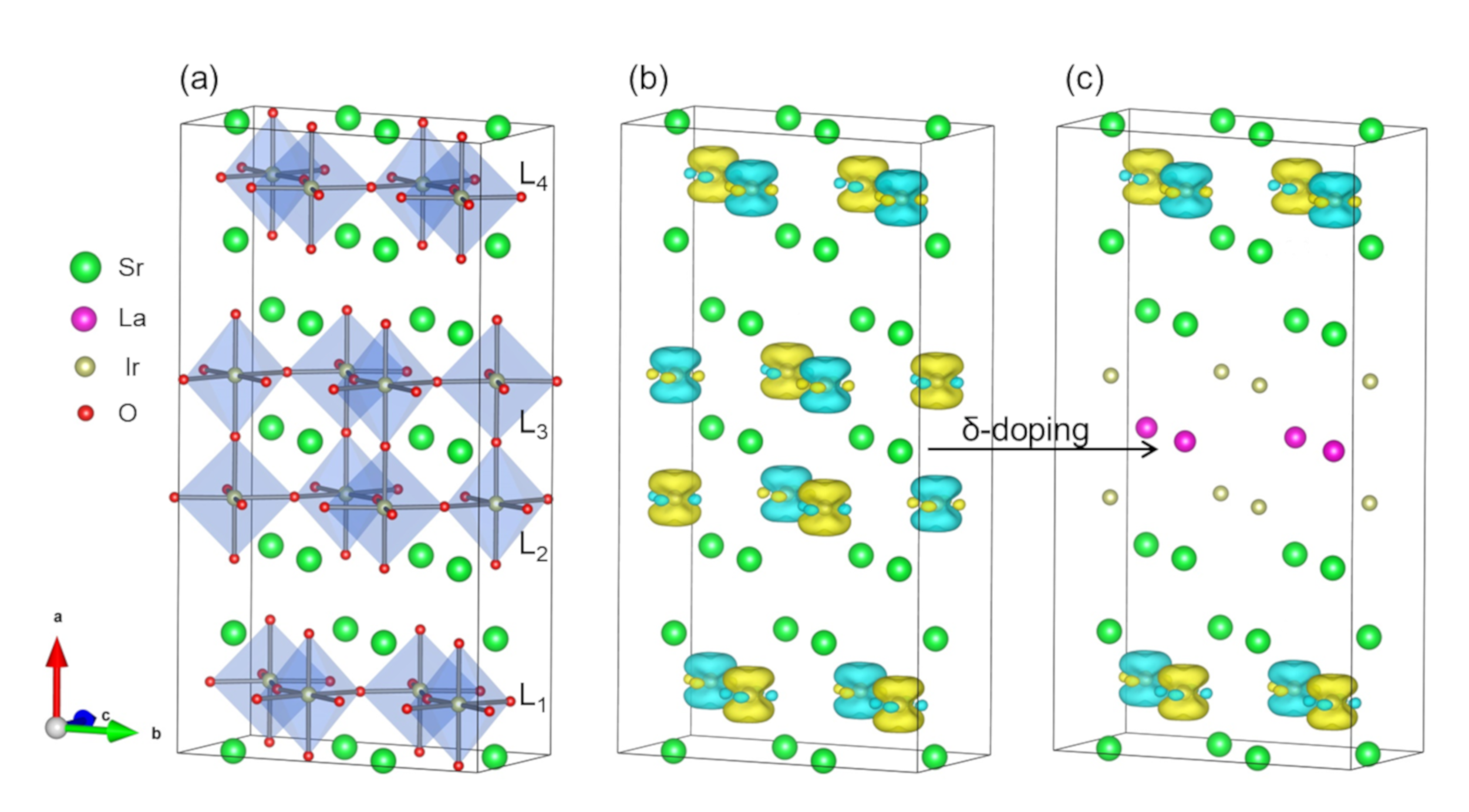}
    \caption{(a) The crystal structure of SIO-327. (b) The spin density contours depicting the G-type magnetic ordering of pristine SIO-327. The yellow and aqua colors represent spin-up and spin-down weights, respectively. The isolevel corresponds to the spin density of 0.004 e$^-$/\AA$^3$. (c) The spin density contours for $\delta$-doped SIO-327, where a SrO layer is replaced with LaO layer. The G-type magnetic order gets destroyed due to electron doping of half-filled $J_{eff}$ = 1/2 state, leading to non-magnetic behavior of L$_2$ and L$_3$ layers, respectively. The IrO$_2$ layers away from the interface, i.e., L$_1$ and L$_4$, continue to show insulating behavior with bulk magnetic ordering.}
    \label{spin-density}
\end{figure*}

Motivated by the aforementioned observations, in this work, we pursue density functional theory (DFT) calculations on $\delta$-doped SIO-327, where a single SrO layer is replaced by LaO layer. A single SIO-327 has six SrO layers along [100]. Therefore, to construct a $\delta$-doped structure, it is sufficient to consider one unit cell and replace one of the SrO layers by LaO layer as shown in Fig. \ref{spin-density}. The periodic superlattice formed out of it introduces a separation of 21 \AA between two consecutive LaO layers which is reasonably large to ignore any interactions between these two layers. 

We predict the formation of a two-dimensional spin-orbital entangled electron gas in $\delta$-doped SIO-327. The 2DEG is found to be non-spin polarized and sharply confined on two IrO$_2$ layers close to the LaO layer. While the extra La valence electron is bound along crystal $a$ direction due to strong confinement potential formed due to the bilayer crystal structure, it is highly mobile in the $bc$ plane. The spin density analysis suggests that the N\'eel state of pristine SIO-327 gets destroyed for those IrO$_2$ layers which are adjacent to the LaO layer. The IrO$_2$ layers far away from the interface remain insulating and preserve the G-type magnetic ordering of pristine SIO-327. Moreover, the tight-binding model developed with the t$_{2g}$ orbitals basis very well captures the essential features of the electronic structure obtained from DFT such as the bulk gap and partially filled electron bands forming the 2DEG. The conductivity tensors estimated using semiclassical Boltzmann theory at room temperature infers that the 2DEG possess ultra-high planar conductivity tensors $\sigma_{yy,zz}$ ($\approx$ 10$^{19}$) which are three-orders higher as compared to normal tensor $\sigma_{xx}$, manifesting the formation of 2DEG.

\section{Structural and Computational details}
The bilayer crystal structure of SIO-327 is shown in Fig. \ref{spin-density}. It crystallizes in C2/c space group with monoclinic crystal structure. The IrO$_6$ octahedra are distorted, the in-plane rotation angle is $\approx 11.81 ^\circ$ and out-of-plane tilt angle is $\approx 0.32 ^\circ$, respectively. For pristine SIO-327, the DFT+$U$+SOC calculations were carried out on the experimental structure \cite{Hogan2015} whereas for $\delta$-doped structure, the calculations were performed on the optimized structure obtained by relaxing the atomic positions and cell volume while keeping the space group symmetry intact. All calculations were performed on a 1$\times$2$\times$1 supercell which is sufficient enough to accomodate the G-type magnetic structure of SIO-327. The strong correlation effect was incorporated via an effective onsite correlation parameter $U_\mathrm{eff}$ = $U-J$ = 2 eV through the rotationally invariant approach introduced by Dudarev \cite{Dudarev1998}. This choice of $U$ is reasonable for SIO-327 as the experimentally reported value of band gap ($\approx$ 0.27 eV) \cite{Swift2018, king2013} is in close agreement with the theoretically estimated one at $U_{eff}$ = 2 eV. The plane-wave based projector augmented wave method (PAW) \cite{Bloch1994,Kresse1999} was utilized to perform DFT calculations in Vienna ab-initio simulation package (VASP) \cite{Kresse1996} within the Perdew$-$Burke$-$Ernzerhof generalized gradient approximation (PBE-GGA) for exchange-correlation functional. The Brillouin zone integrations were carried out using $1 \times 4 \times 8$ $\Gamma$-centered $k$-mesh. The kinetic energy cutoff for plane-wave basis set was chosen to be $400$ eV. The planar and macroscopic average potentials were calculated using QUANTUM ESPRESSO \cite{Giannozzi2009}. The principal components of conductivity tensors $\sigma_{\alpha\beta}$ were computed at room temperature by employing semiclassical Boltzmann transport theory as implemented in VASPKIT \cite{VASPKIT}. A 5$\times$18$\times$18 $k$-mesh was used to obtain the smooth interpolation of bands and to compute the necessary
derivatives which were required for the calculation of $\sigma_{\alpha\beta}$.
\section{Bulk electronic structure}
\begin{figure}
    \centering
    \includegraphics[angle=-0.0,origin=c,height=3.1cm,width=8.45cm]{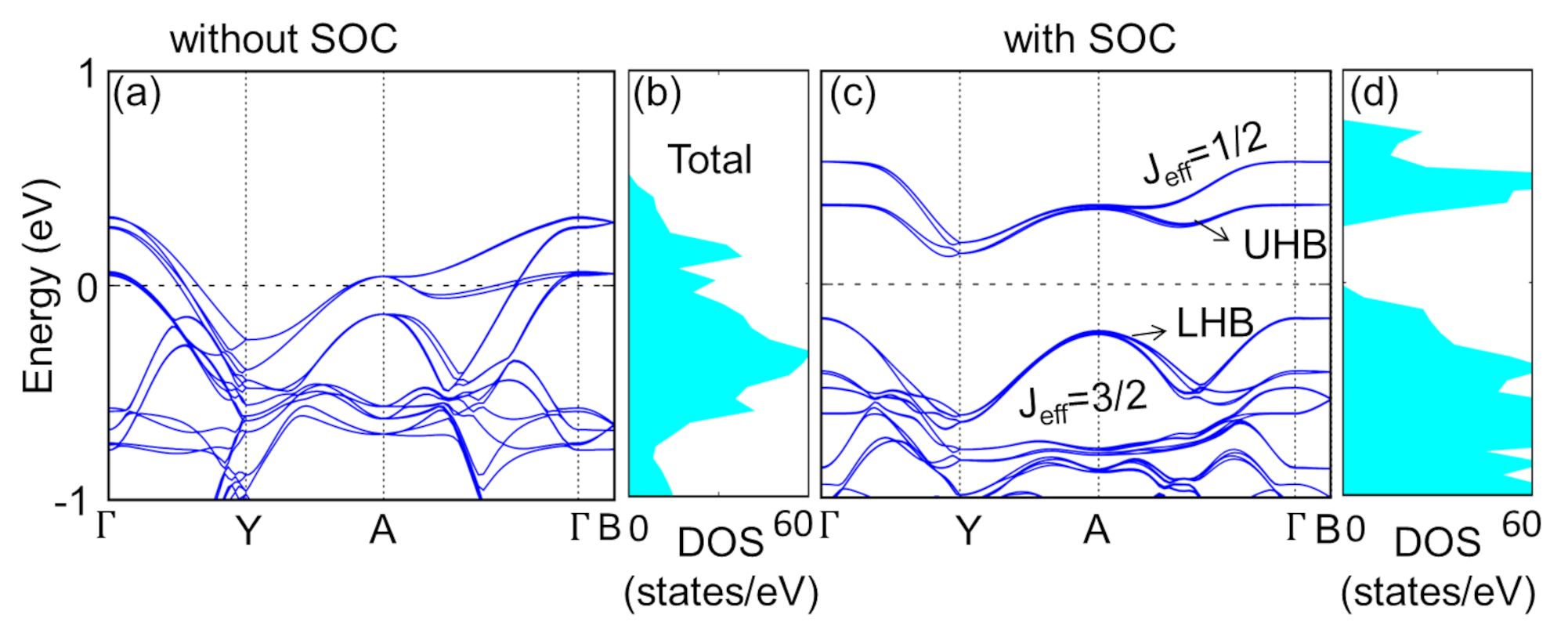}
    \caption{(a,b) The band structures and total density of states of SIO-327 within DFT+$U$ and (c,d) within DFT+$U$+SOC with $U$ = 2 eV. The $t_{2g}$ states split to form completely occupied $J_{eff}$ = 3/2 states and half-filled $J_{eff}$ = 1/2 state. The opening of a gap with the inclusion of SOC manifests the spin-orbit-assisted Mott insulating ground state. The dashed line represents the Fermi level.}
    \label{bulk-band}
\end{figure}

Before examining the effect of $\delta$-doping on the electronic and magnetic structure of SIO-327, it is prudent to first analyze the ground state electronic and magnetic properties of undoped SIO-327. The Fig. \ref{bulk-band} depicts the band structures and corresponding atom resolved density of states (DOS) within DFT+$U$ (first column) and DFT+$U$+SOC (second column) with the onsite Coulomb repulsion $U_{eff}$ = 2 eV as appropriate for SIO-327.

As shown in Figs. \ref{bulk-band}(a,b), the Fermi level ($E_F$) is well populated by $t_{2g}$ states whereas the $e_g$ states lie far above the $E_F$ and are unoccupied. The latter occurs due to crystal field of IrO$_6$ octahedral complex which split five-fold degenerate $d$ states into unoccupied $e_{g}$ doublet and a low-energy lying and partially occupied $t_{2g}$ triplet (see Fig. \ref{split-d-states}). The latter leads to a metallic state even with the inclusion of onsite Coulomb repulsion. With the inclusion of SOC along with $U$, see Figs. \ref{bulk-band}(c,d), the $t_{2g}$ states further split to form spin-orbital entangled states which are expressed as
\begin{align}
  \bigg| \frac{1}{2}, \pm{\frac{1}{2}} \bigg \rangle &= \frac{1}{\sqrt{3}}( \ket{yz,\bar{\sigma}} \pm{\ket{xy,\sigma}} \pm{i} \ket{xz,\bar{\sigma}}),\notag\\
   \bigg|\frac{3}{2}, \pm{\frac{1}{2}}\bigg \rangle &= \frac{1}{\sqrt{6}}(\ket{yz,\bar{\sigma}} \mp 2{\ket{xy,\sigma} } \pm{i} \ket{xz,\bar{\sigma}}),\notag\\
    \bigg|\frac{3}{2}, \pm{\frac{3}{2}}\bigg \rangle &= \frac{1}{\sqrt{2}}(\ket{yz,\sigma} \pm{i} \ket{xz,{\sigma}}),
  \end{align}
where $\pm$ corresponds to spin $\sigma$ = $\uparrow$/$\downarrow$, respectively.
The $J_{eff}$ = 3/2 states get completely occupied due to the $d^5$ valence state of Ir, whereas $J_{eff}$ = 1/2 state is half-occupied and splits into LHB and UHB with a narrow gap in between (see Fig. \ref{split-d-states}). The opening of a gap with the inclusion of SOC infers that SIO-327 is a weakly correlated spin-orbit-assisted Mott insulator. This insulating state possesses G-type magnetic ordering (nearest-neighbor spins are antiferromagnetically coupled) which is depicted through spin density shown in Fig. \ref{spin-density}(b). Moreover, the dominant spin-up or spin-down mixture in local spin density reflects the spin mixture of the $\ket{1/2,\pm 1/2}$ wave functions, respectively.
\section{Formation of spin-orbital entangled electron gas}
\begin{figure}
    \centering
    \includegraphics[angle=-0.0,origin=c,height=11cm,width=8.7cm]{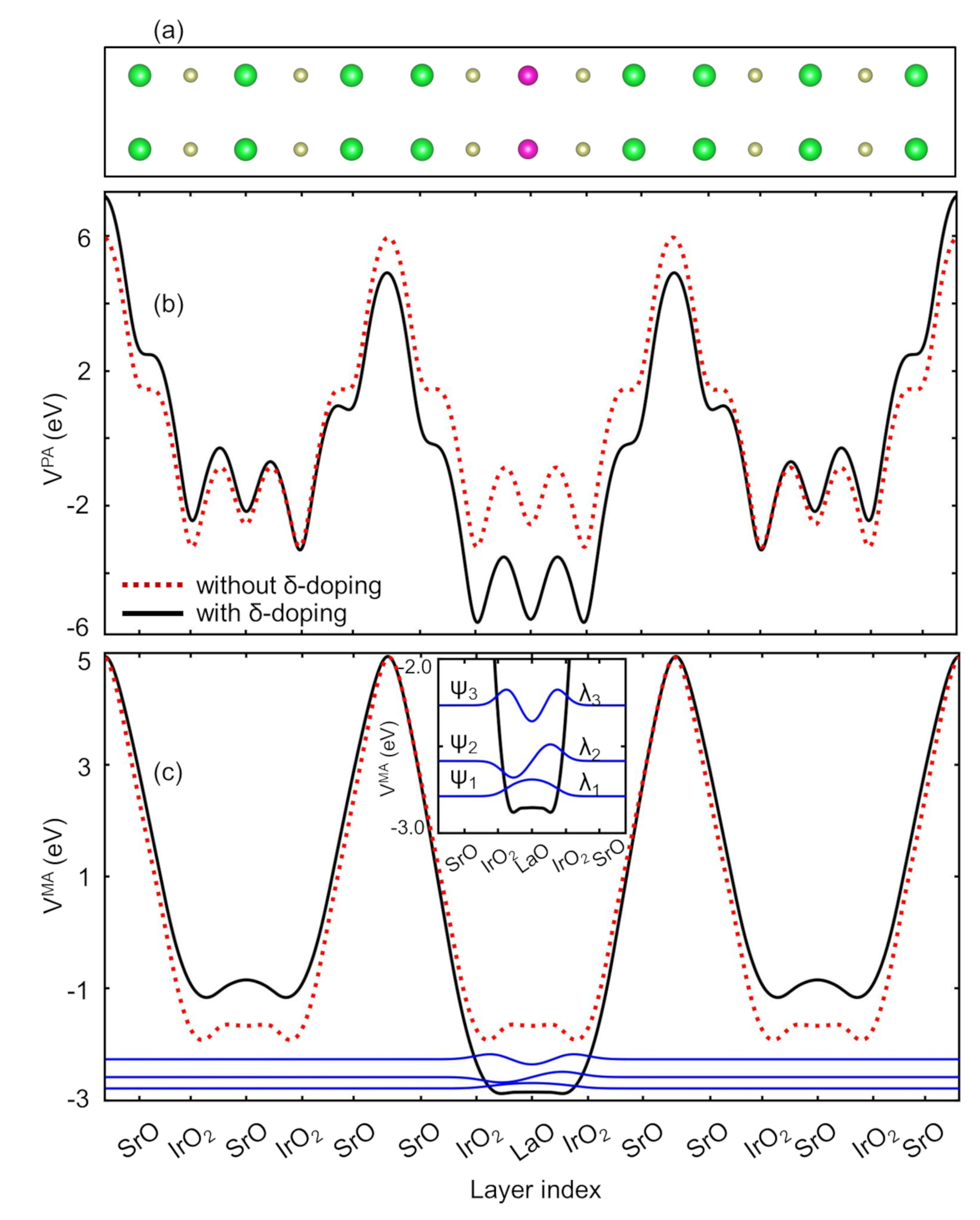}
    \caption{The cell-averaged bare ionic potential along crystal $a$ axis without (red dashed line) and with (black dashed line) $\delta$-doping. The inset represents the ground and the first two excited states wave functions  which are obtained by numerically solving the Schr\"{o}dinger equation in the $\delta$-doped potential.}
    \label{bare-pot}
\end{figure}
The unoccupied spin-orbital entangled state of the electron, i.e., the UHB of Ir, can give rise to an interesting quantum phenomenon upon electron doping. Recent experimental and theoretical studies have reported many intriguing features such as CDW instability \cite{Chu2017}, negative electron compressibility \cite{He2015}, and collapse of the band gap in La substituted SIO-327 \cite{Hogan2015}. To the best of our knowledge, a detailed theoretical analysis of the mechanisms governing the band gap collapse and emergent properties in electron-doped SIO-327 is still lacking in the literature. This motivates us to pursue a detailed electron structure analysis of electron doping in this quasi-bilayer compound. For this purpose, we have performed ab-initio calculations on delta-doped SIO-327 (La$_\delta$SIO-327). As shown in Fig. \ref{spin-density} (c), the La$_\delta$SIO-327 configuration is achieved by replacing a single SrO layer by LaO layer in one of the bilayer units of the pristine crystal structure. Since La has one extra valence electron as compared to Sr, the $\delta$-doping in this work implies the case of electron doping in SIO-327. 

As La is an electron donor, it will donate the extra electron to the system. In $\delta$-doped 3$d$ perovskite oxide SrTiO$_3$ \cite{Matsubara2016}, it was found that the extra La electron spreads upto several TiO$_2$ layers around LaO which makes the Fermi surface complex as it is populated by many Ti-$d$ states. To analyze the spread of electrons in La$_\delta$SIO-327, we have calculated the variation of the planar and macroscopic cell-average of the electrostatic potential $V^{PA}$ and $V^{MA}$ as a function of layers without (red dashed line) and with (black solid line) considering $\delta$-doping. The $V^{MA}$ is calculated in two steps. In the first step, the raw three-dimensional potential $V^{raw}$ is averaged in the $yz$ plane to obtain planar-average potential $V^{PA}$: 
\begin{equation}
V^{PA}(x)=\frac{1}{S}\int V^{raw}(x,y,z)dydz,
\end{equation}
where $S$ is the area of [100] plane of the unit cell. In the second step, the $V^{PA}$ is averaged further over a period $c$ normal to the $bc$ plane to obtain $V^{MA}$:
\begin{equation}
    V^{MA}(x)=\frac{1}{c}\int_{x-c/2}^{x+c/2} V^{PA}(x)dx,
\end{equation}
where $c$ is the length at which averaging is performed. The $c$ is chosen to be $\approx$ 3.6 \AA which is the SrO-IrO inter-layer distance. The calculated $V^{PA}$, $V^{MA}$, and the electron wave functions for ground and first two excited states are shown in Fig. \ref{bare-pot}.

\begin{figure}
    \centering    \includegraphics[angle=-0.0,origin=c,height=10.5cm,width=8.2cm]{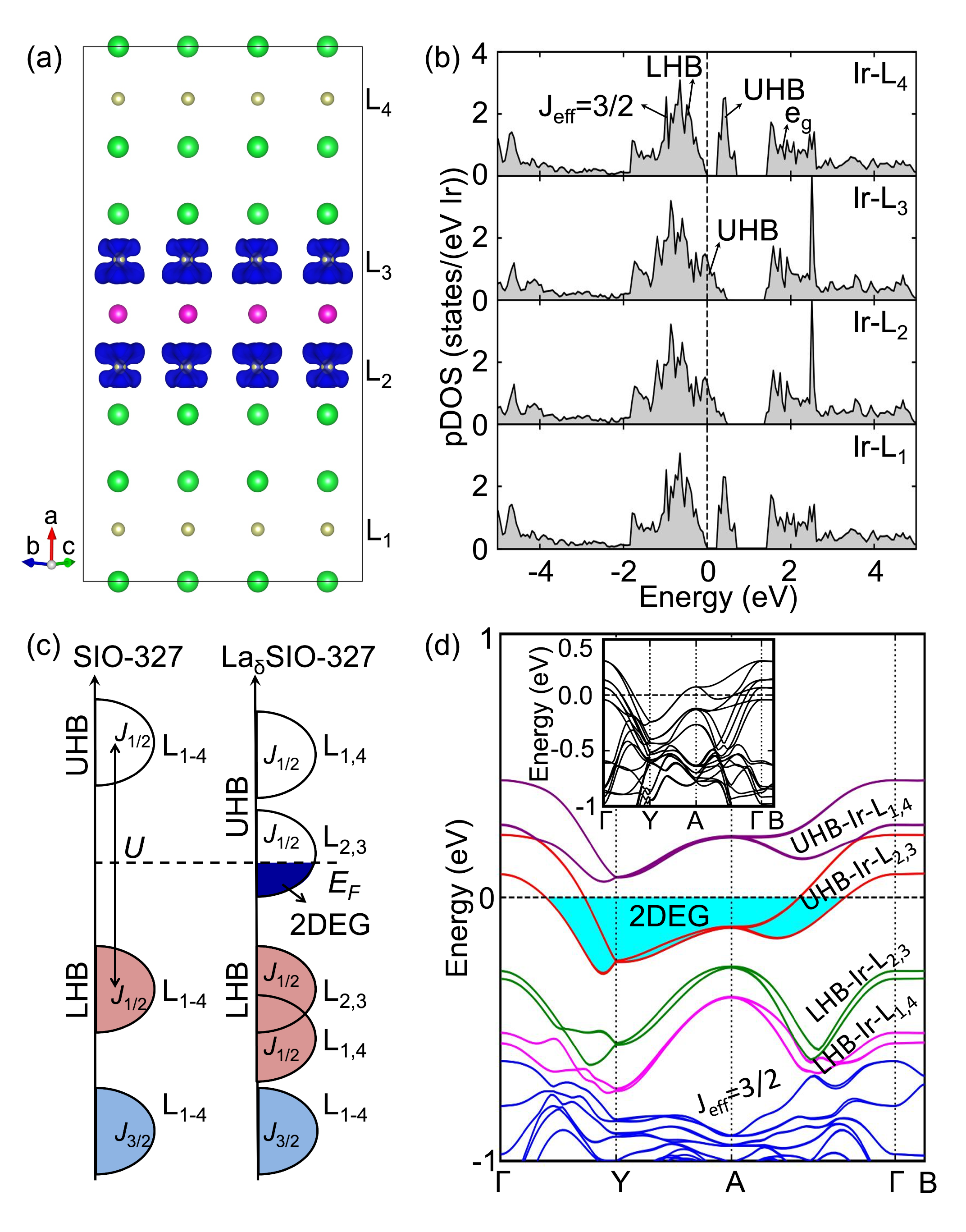}
    \caption{(a) The charge density contours for $\delta$-doped SIO-327. The charge spread is restricted to Ir atoms of layers L$_2$ and L$_3$. The isolevel corresponds to the charge density of 0.004 e$^-$/\AA$^3$. The oxygen atoms are not shown for clarity. (b) The layer resolved Ir densities-of-states. While the Mott state of the IrO$_2$ adjacent to the LaO layer destroys to make layers L$_2$ and L$_3$ metallic, the layers far away from the interface, i.e., layers L$_1$ and L$_4$, remain insulating. (c,d) The schematic illustration of the electronic states with and without $\delta$-doping and the corresponding band structure of La$_\delta$SIO-327. The UHB of Ir atoms in layers L$_2$ and L$_3$ gets partially occupied to host 2DEG. The inset represents the band structure of La$_\delta$SIO-327 without SOC. The 3$\lambda$/2 gap between $J_{eff}$ = 3/2 and $J_{eff}$ = 1/2 bands is absent due to the absence of SOC.}
    \label{whole}
\end{figure}

Even without $\delta$-doping, a potential well forms along crystal $a$ direction with depth $\approx$ 6 eV. This confinement potential exists even in the bulk system due to the quasi-two-dimensional nature of SIO-327. As a consequence, different layers exhibit uneven potential which averages out to form periodic quantum wells (see Figs. \ref{bare-pot} (b,c)). On the contrary, in a regular perovskite structure, such confinement potential will vanish due to the three-dimensional nature of the crystal structure. With $\delta$-doping, the well depth gets further modulated asymmetrically as the LaO layer repeats after lattice period $a$. Due to the large potential barrier, presumably, the extra La electron will spread up to two IrO$_2$ layers on either side of the LaO, i.e., layers L$_2$ and L$_3$. To validate it, we plot the charge density contours which are shown in Fig. \ref{whole} (a). These contours are obtained by subtracting the charge densities of un-doped and doped SIO-327. As expected, while the Ir atoms corresponding to layers L$_2$ and L$_3$ hold most of the charge, the Ir atoms far away from the $\delta$-doped layer do not gain any charge. The inset of Fig. \ref{bare-pot} (c) shows the wave functions (blue solid lines) corresponding to the ground and first two excited states ($\psi_n(r)$ with $n$ = 1, 2, 3,...) in $\delta$-doped potential. These are obtained from the numerical solution to one particle Schr\"{o}dinger equation. The strong localization of the wave functions inside the well and rapid decay outside the well further validate the obtained charge contours. Moreover, these wave functions resemble to the eigen solutions of the one-dimensional harmonic oscillator.
\begin{table}
\caption{The estimated spin/orbital magnetic moments and distribution of $\delta$-doped electron, one per La atom, among Ir atoms of various layers in La$_\delta$SIO-327.}
\begin{center}
\begin{tabular}{|c|c|c|c|}
\hline
atom-layer & $m_s$ ($\mu_B$) & $m_l$($\mu_B$) & charge (e$^-$)\\ 
\hline
Ir-L$_1$ & 0.26 & 0.35 & 0.00\\
Ir-L$_2$ & 0.00 & 0.00 & 0.35\\
Ir-L$_3$ & 0.00 & 0.00 & 0.35\\
Ir-L$_4$ & 0.26 & 0.35 & 0.00\\
\hline
\end{tabular}
\end{center}
\label{spin-orbital-charge}
\end{table}

It is important to note that, in $\delta$-doped strongly correlated 3$d$ transition metal oxide SrTiO$_3$, \cite{Matsubara2016} the 2DEG was found to spread up to several TiO$_2$ layers. In our study, the 2DEG is found to be confined in only two IrO$_2$ layers and is spin-orbital entangled. Such confinement of 2DEG in few layers makes the Fermi surface devoid of large DOS. Therefore, such systems, once experimentally synthesized, will possess easier tunability and can be engineered to produce emergent quantum phases.

To estimate the charged gain by Ir atom of layers L$_2$ and L$_3$ and to examine the electronic and magnetic structure of La$_\delta$SIO-327, in Figs. \ref{whole} (b,c,d), we have plotted the layer resolved DOS, the schematic representation of the electronic states, and the band structure, respectively. The primary analysis of the ideal charge gained by Ir atoms can be made by analyzing the nominal charge states of single formulae unit of doped and undoped systems. In the undoped system Sr$^{6+}_3$Ir$^{8+}_2$O$_7^{14-}$, Sr and O ions possess 2+ and 2- charge states which give rise to Ir$^{4+}$ charge state and hence $d^5$ electronic configuration. As La possesses 3+ charge state, the adjacent IrO$_2$ layers on both the sides of LaO layer gain one additional electron to make the nominal charge state of Ir as 3.5. This is ideal when the electron gained is confined to the adjacent IrO$_2$ layers. The DFT calculations which are discussed next will let us know how much deviation occurs from this ideal distribution. 

To quantify the distribution of $\delta$-doped electron (one per La atom), we integrate the DOS for UHB of Ir atoms in layers L$_2$ and L$_3$ from $E_F$ to conduction band top. The obtained charges are listed in Table \ref{spin-orbital-charge}. As expected, while the dominant charge is held by Ir atoms ($\approx$ 70\%), the rest is distributed among other atoms.

As inferred from band structure, the $\delta$-doping leads to the formation of a metallic state in SIO-327. Due to charge spread, the Ir atom in layers L$_2$ and L$_3$ gets electron-doped (see charge contours in Fig. \ref{whole}). Since, in bulk SIO-327, Ir exhibits half-filled $J_{eff}$ = 1/2 state due to completely occupied $J_{eff}$ = 3/2 states, which are now electron-doped, the antiferromagnetic Mott insulating state gets destroyed (see the spin and orbital polarization in Table \ref{spin-orbital-charge}). As a result, the UHB of the Ir atom from layers L$_2$ and L$_3$ gets partially occupied to form non-spin polarized 2DEG. However, the UHB of Ir atoms corresponding to layers L$_1$ and L$_2$ remains unoccupied and continues to show bulk insulating behavior with G-type magnetic ordering (see spin density in Fig. \ref{spin-density} (c)). We did not notice any charge ordering which could have made the system insulator with the stabilization of 3+ and 4+ charge states of Ir, respectively. 

The layer resolved Ir DOS shown in Fig. \ref{whole}(b) further confirms the picture made through band structure analysis. As can be seen clearly, layers L$_1$ and L$_4$ are insulating, whereas layers L$_2$ and L$_3$ possess finite DOS at the $E_F$ leading to metallicity and collapse of magnetic ordering. 

\begin{table*}
\caption{The estimated values of on-site energy and tight-binding parameters in eV units. The $\epsilon$ is the onsite energy of Ir ions in the pristine SIO-327. V$_1$ and V$_2$ are the additional potential energies on Ir ions in layers $L_{1,4}$ and $L_{2,3}$, respectively because of $\delta$ doping of La. The $t_1$ and $t_2$ are nearest-neighbor (NN) intra-layer and next-nearest-neighbor (NNN) inter-layer $\sigma$ interactions, whereas, $t_3$/$t_4$ and $t_5$/$t_6$ denote $\pi$ interactions for NN/NNN intra-layer and inter-layer hopping interactions, respectively.}
\begin{center}
\begin{tabular}{c c c c c c c c c c}
\hline
\hline
system & $\epsilon$ & V$_1$ & V$_2$ & $t_1$ & $t_2$ & $t_3$ & $t_4$ & $t_5$ & $t_6$\\ 
\hline
SIO-327& $-5.787$ & $0$ & $0$ & $-0.267$ & $0.021$ & $0.227$ & $-0.014$ & $-0.038$ & $-0.008$\\
La$_\delta$SIO-327 & $-5.787$ & $0$ & $-1.36$ & $-0.267$ & $-0.01$ & $0.227$ & $-0.014$ & $0.013$ & $0.001$\\ 
\hline
\hline
\end{tabular}
\end{center}
\label{doped-tb}
\end{table*}
\section{Model Hamiltonian}
\begin{figure}
    \centering    \includegraphics[angle=-0.0,origin=c,height=4.3cm,width=8.6cm]{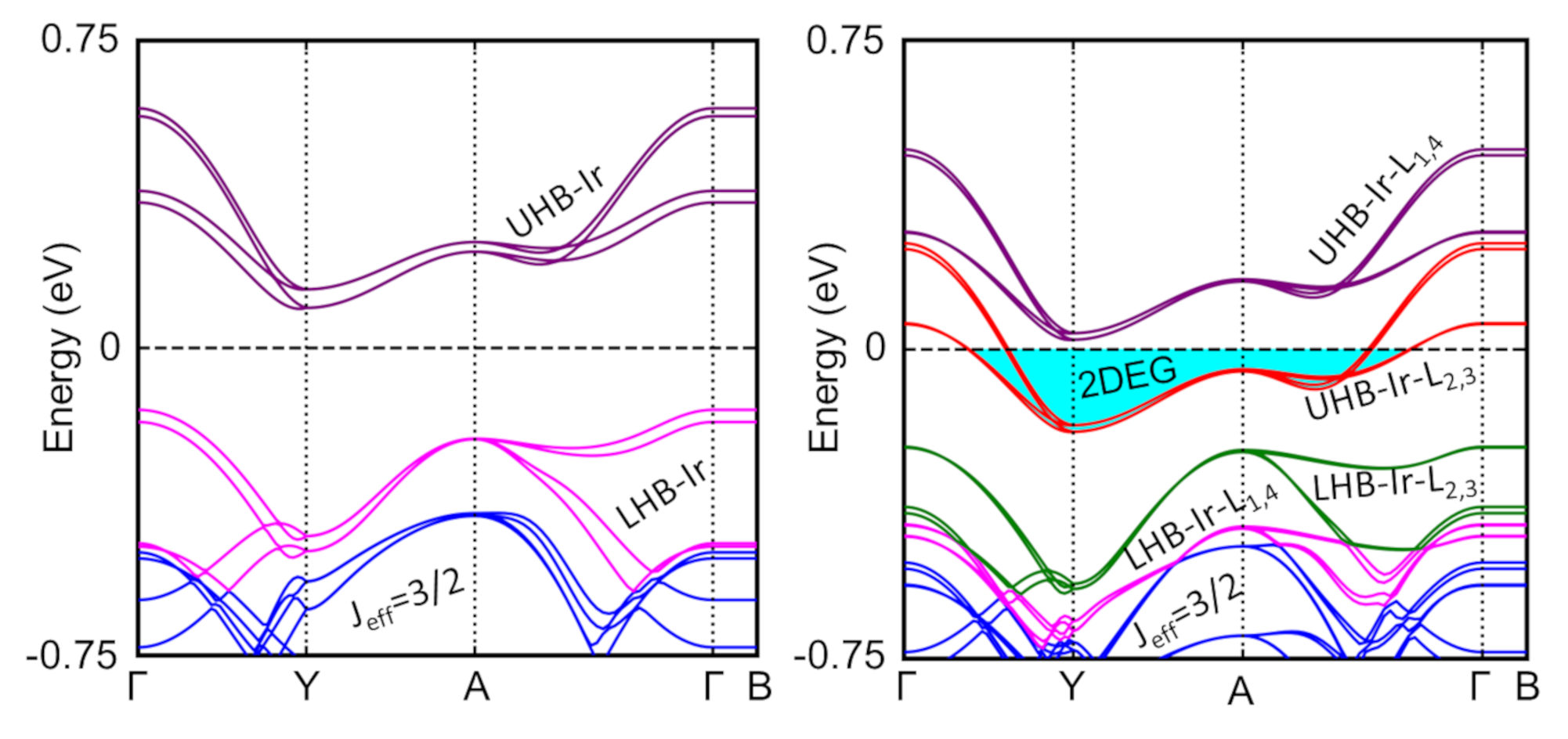}
    \caption{The band structures of undoped (left) and doped SIO-327 (right) as obtained by solving the tight-binding model Hamiltonian (Eq. \ref{MOH}).}
    \label{tb}
\end{figure}
To gain more insights into the formation of spin-orbital entangled 2DEG, we have developed a $t_{2g}$ orbitals based tight-binding (TB) model which also incorporates SOC, electron-electron correlation terms, and the effect of the potential well. This  multi-orbital Hubbard-Kanamori Hamiltonian is given by
 \begin{equation} \label{MOH}
  \begin{split}
    H & = \sum_{i, \mu} (\epsilon_{i\mu} + V_{i}) c_{i\mu}^\dagger c_{i\mu}+ \sum_{i, j, \mu, \nu} (t_{i\mu j\nu} c_{i\mu}^\dagger c_{j\nu} + h.c.) + \\ 
    & U \sum_{i, \mu} n_\mathrm{i\mu\uparrow} n_\mathrm{i\mu\downarrow} + (U^\prime -\frac{J_\mathrm{H}}{2}) \sum_{i, \mu < \nu} n_\mathrm{i\mu} n_\mathrm{i\nu} \\ 
    & -2J_\mathrm{H} \sum_{i, \mu < \nu} S^z_\mathrm{i\mu} \cdot S^z_\mathrm{i\nu}+\lambda \sum_{\alpha, \beta, \sigma, \bar{\sigma}} \bra{\alpha \sigma} \boldsymbol{L} \cdot \boldsymbol{S} \ket{\beta \bar{\sigma}} c_{\alpha, \sigma}^\dagger c_{\beta,\bar{\sigma}},
\end{split}
\end{equation}
where, $i$ ($j$) and $\mu$ ($\nu$) represent sites and orbitals indices, respectively. The first two terms in $H$ describe the onsite and the kinetic energy of the electrons, while the third and fourth terms represent the energy cost of putting the electrons in the same and different orbitals. The last two terms define Hund's rule and spin-orbit coupling. In the first term, $\epsilon_{i\mu}$ is the onsite energy of Ir-t$_{2g}$ orbitals for the pristine SIO-327, and $V_{i}$ is the additional potential energy realized by the Ir ions due to $\delta$-doping of La. In the fourth term, the relation $U^\prime$ = $U$ - $2J_\mathrm{H}$ has been used. The $J_H$/$U$ ratio is taken to be 0.2 with $U$ = 2 eV, which is found to be appropriate for 4$d$/5$d$ oxides \cite{Meetei2015}. The SOC strength $\lambda$ of Ir atom is fixed to 0.43 eV, as obtained by estimating the gap between $J_{eff}$ = 3/2 and $J_{eff}$ = 1/2 bands (see Fig. \ref{split-d-states}). The $n_\mathrm{i\mu}$ = $n_\mathrm{i\mu\uparrow}$ + $n_\mathrm{i\mu\downarrow}$ are the occupation numbers, which are calculated from the DFT+$U$  density matrix.

 We first develop a TB Hamiltonian for the undoped SIO-327. 
 Later, we introduce the additional potential energy $V_{i}$ obtained from the bare macroscopic potential of undoped and doped SIO-327 (see Fig. \ref{bare-pot}), to the Ir ions in layers L$_2$/L$_3$. The numerical values of V$_i$ can be found in Table \ref{doped-tb}. Because of $\delta$ doping, one extra electron comes from each La atom, which is uniformly distributed among Ir ions in layer L$_2$/L$_3$ in the TB model. The TB parameters for the undoped and doped cases are listed in Table \ref{doped-tb}, which are obtained by fitting TB bands with DFT. The corresponding band structures are shown in Fig. \ref{tb}. The TB bands very well capture the essential features of the DFT bands, as shown in Figs. \ref{bulk-band}(c) and \ref{whole}(d), such as the bulk band gap and the partially filled electron bands forming the 2DEG.

\section{Transport Properties}
\begin{figure}
    \centering
    \includegraphics[angle=-0.0,origin=c,height=4.9cm,width=8.6cm]{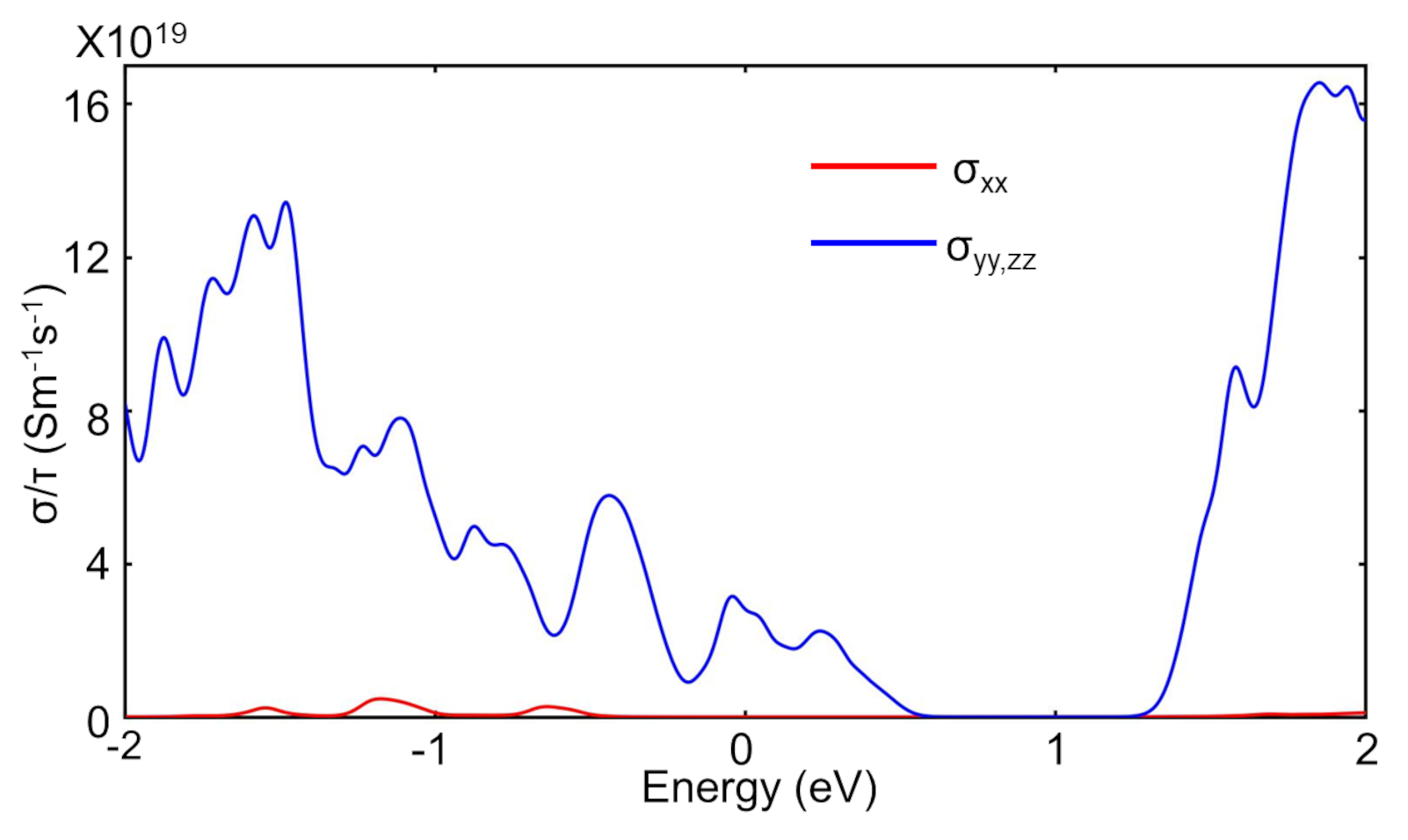}
    \caption{The transport properties of $\delta$-doped SIO-327. The principal components of the electrical conductivity tensor at room temperature obtained using semi-classical
Boltzmann transport theory. The confinement potential restricts the electron motion along [100] direction, and hence, $\sigma_{xx}$ becomes negligible. Significant orders of difference between $\sigma_{yy,zz}$ and   $\sigma_{xx}$ imply two-dimensional mobility of the electrons.}
    \label{conductivity}
\end{figure}
The formation of spin-orbital entangled 2DEG through the confinement effect can be quantified by calculating the conductivity. For this purpose, we have adopted semi-classical Boltzmann transport theory and calculated the conductivity tensors $\sigma$ from the first-order derivative of the bands $\epsilon$(k):
\begin{equation}
    \sigma_{\alpha\beta}(\epsilon) = \frac{e^2\tau}{N}\sum_{i,k}v_\alpha(i,k) v_\beta(i,k) \frac{\delta(\epsilon-\epsilon_{i,k})}{d\epsilon},
\end{equation}
where $\tau$ is the relaxation time, $i$ is the band index, $v$ is the first-order derivative of $\epsilon_{i,k}$ and $N$ is the number of $k$ points sampled. The notations $\alpha$ and $\beta$ denotes the crystal axes. The temperature-dependent conductivity evaluated using Eq. 4 is given by
\begin{equation}
    \sigma_{\alpha\beta}(T,\mu) = \frac{1}{\Omega}\int \sigma_{\alpha \beta}(\epsilon)[-\frac{\partial{f_\mu(T,\epsilon)}}{\partial{\epsilon}}] \textit{d}\epsilon,
\end{equation}
where $\Omega$ is the volume of the unit cell, $\mu$ (= $E_F$) is the chemical potential, and $f$ is the Fermi-Dirac distribution function. The estimated conductivity tensors are shown in Fig. \ref{conductivity} where we have plotted $\sigma$/$\tau$ vs Energy at room temperature. Due to confinement potential, while the electron motion is bound along the $x$ direction, in the $yz$ plane, it is free. This is very well reflected in the conductivity tensors. The conductivity tensor $\sigma_{xx}$ is negligible as compared to $\sigma_{yy,zz}$. At $E_F$, the $\sigma_{yy,zz}$ is four-order higher as compared to $\sigma_{xx}$. Interestingly, the 2DEG formed out of partially occupied Ir UHB possess ultra-high conductivity of the order of $\approx$ 10$^{19}$ Sm$^{-1}$s$^{-1}$.

\section{Summary}
To summarize, by pursuing density-functional studies and a relevant tight-binding model, we have examined the electronic structure of $\delta$-doped quasi-two-dimensional iridate Sr$_3$Ir$_2$O$_7$ (SIO-327), where a single SrO layer is replaced by LaO layer, we predict the formation of a two-dimensional spin-orbital entangled electron gas (2DEG). A strong confinement potential forms along the $x$ direction due to the quasi-two-dimensional nature of SIO-327, as well as due to the presence of positively charged LaO layer, the extra La electron gets bound and is highly mobile in the $bc$ plane. The charge analysis suggests that nearly 70\% of the doped electrons are confined to the IrO$_2$ planes adjacent to the LaO layer. As a consequence, the half-filled $J_{eff}$ = 1/2 state gets electron-doped, leading to the destruction of the antiferromagnetic Mott insulating state and partially occupied Ir upper-Hubbard subbands that host the 2DEG. The IrO$_2$ layers away from the interface preserve the G-type magnetic ordering of pristine SIO-327. The conductivity tensors calculated using semi-classical Boltzmann theory reveal that the 2DEG possess ultra-high planar conductivity tensors $\sigma_{yy,zz}$ ($\approx$ 10$^{19}$), which are four-orders higher as compared to normal tensor $\sigma_{xx}$. Our study will encourage the experimenters to grow $\delta$-doped structures for a wide class of spin-orbit correlated materials to explore the formation and application of spin-orbital entangled 2DEG.\\

\section{Acknowledgement}
The authors would like to thank HPCE, IIT Madras for providing the computational facility. This work is funded by the Department of Science and Technology, India, through grant No. CRG/2020/004330.

\end{document}